\documentclass{ws-procs9x6}
\usepackage{unites2e}
\begin{document}

\title{VERITAS: Status and Performance}

\author{J. Holder$^*$ for the VERITAS collaboration}

\address{Department of Physics and Astronomy, University of Delaware,\\
Newark, DE 19716, USA\\
$^*$E-mail: jholder@physics.udel.edu}

\begin{abstract}
VERITAS is an atmospheric Cherenkov telescope array sited in Tucson, Arizona. The array is nearing completion and consists of four, $12\U{m}$ diameter telescopes. The first telescope in the array has been operating since February 2005, while observations with the full array are expected to begin in January, 2007. We report here in some detail on the performance of the first VERITAS telescope, and briefly discuss the first stereo observations.
\end{abstract}


\bodymatter

\section{Introduction}
The Whipple 10\U{m} telescope provided the first detection of
an astrophysical $\gamma$-ray source, the Crab Nebula, using
the technique of forming images of the atmospheric Cherenkov emission generated by air showers \cite{Weekes89}. 
Arrays of imaging atmospheric Cherenkov telescopes
provide a further increase in sensitivity and in angular and energy resolution,
as demonstrated by the HEGRA experiment \cite{Puhlhofer03}.
The stereoscopic imaging atmospheric Cherenkov technique is now being exploited by four major atmospheric Cherenkov observatories; HESS\cite{Hinton04}, MAGIC\cite{Lorenz04}, CANGAROO~III\cite{Kubo04} and VERITAS \cite{Weekes02}. The VERITAS array comprises four, 12\U{m} telescopes and will be used to observe astrophysical sources from the northern hemisphere over the energy range from $100\U{GeV}$ to $50\U{TeV}$, with a sensitivity of 7 mCrab (a gamma-ray source of 0.7\% of the Crab Nebula flux will be detected with $5\sigma$ significance over a 50 hour exposure). The first telescope came online in February 2005\cite{Holder06}, followed by the first stereo observations with two telescopes in January 2006. We present here a detailed description of the performance of the first telescope and some initial results from the stereo system.

\section{The First VERITAS Telescope}
\begin{figure}
\begin{center}
\psfig{file=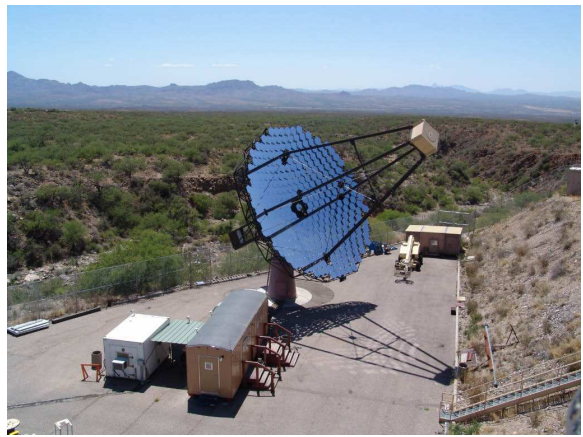,width=3in}
\end{center}
\caption{The first VERITAS telescope installed at the Whipple Observatory basecamp}
\label{T1}
\end{figure}

Figure~\ref{T1} shows the first VERITAS telescope installed at the Whipple observatory basecamp in Tucson, Arizona, at an altitude of $1275\U{m}$ above sea level.
The $12\U{m}$ diameter collector is used to produce an image of the Cherenkov emission generated by gamma-ray and cosmic-ray air showers on a 499 photomultiplier tube (PMT) camera at the focus.
The buildings in the foreground house the trigger and data acquisition electronics and the power supply systems. 

\vspace{-0.3in}
\subsection{Optical and Mechanical Performance}
\begin{figure}
\begin{center}
\psfig{file=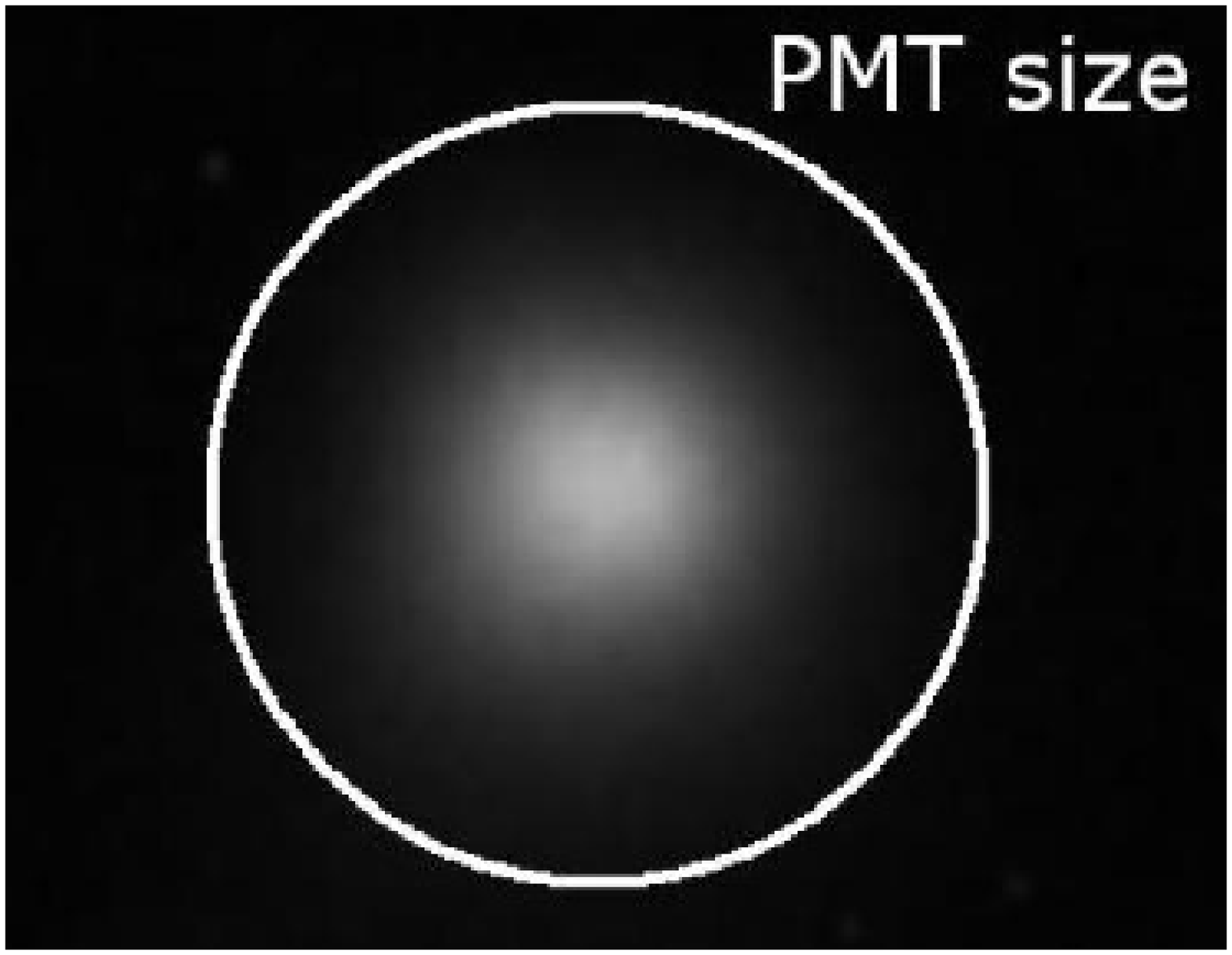,width=2.2in}\hspace{0.15in}\psfig{file=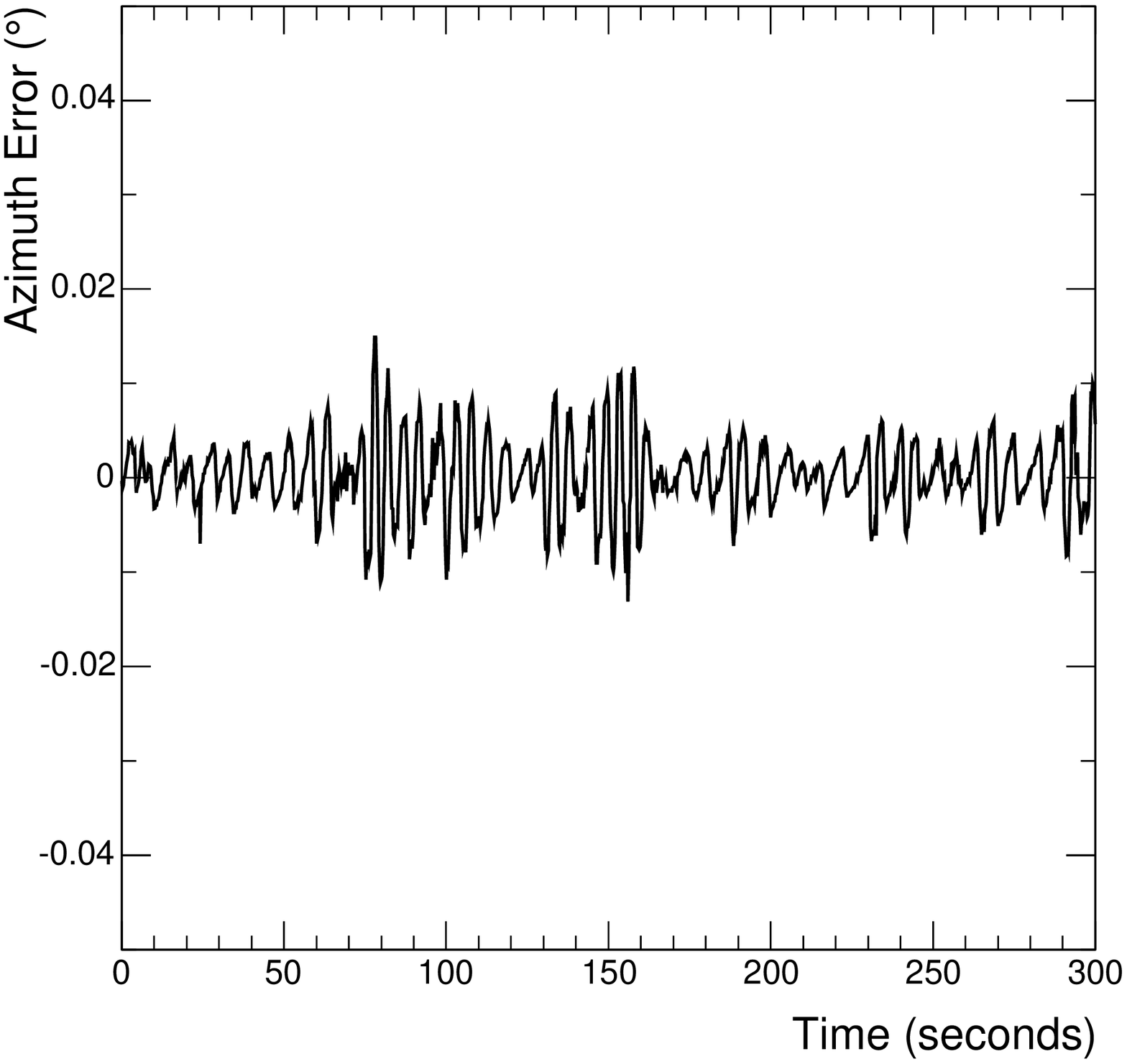,width=1.9in}
\end{center}
\caption{{\bf Left:} An image of Polaris in the focal plane recorded with a CCD camera. The circle indicates the size of a PMT $(0.15^{\circ})$. {\bf Right:} The Azimuth encoder residuals (difference between measured and requested position) for a short tracking run.}
\label{optomech}
\end{figure}

The telescope optics consist of a tessellated reflector of Davies-Cotton design \cite{Davies57} with $12\U{m}$ diameter and a focal length of $12\U{m}$.
The reflector is comprised of 350 individual facets, each with an area of $0.32\UU{m}{2}$, giving a total mirror area of $\sim110\UU{m}{2}$.
The facets are made from glass, which is slumped and polished, then cleaned, aluminized and anodized at an on-site optical coating laboratory. 
The reflectivity of the coating is typically $>90\%$ at $320\U{nm}$.  
Reflector facets are aligned manually using a laser alignment system located at a distance of twice the focal length ($24\U{m}$) from the centre of the reflector.
The point spread function response is shown in Figure~\ref{optomech} and is measured to be $\sim0.06^{\circ}$ (full width at half-maximum) at the position of Polaris (elevation $31^{\circ}$). 

The mirrors are distributed over a tubular steel optical support structure which is mounted on a altitude-over-azimuth positioner. 
The maximum slew speed of the positioner is measured to be $\sim1^{\circ}\UU{s}{-1}$.
Figure~\ref{optomech} shows the encoder measurements for a short tracking run, illustrating that the tracking is stable with a relative pointing accuracy of typically $<\pm0.01^{\circ}$.

\vspace{-0.3in}
\subsection{The Camera and Electronics}
\begin{figure}
\begin{center}
\psfig{file=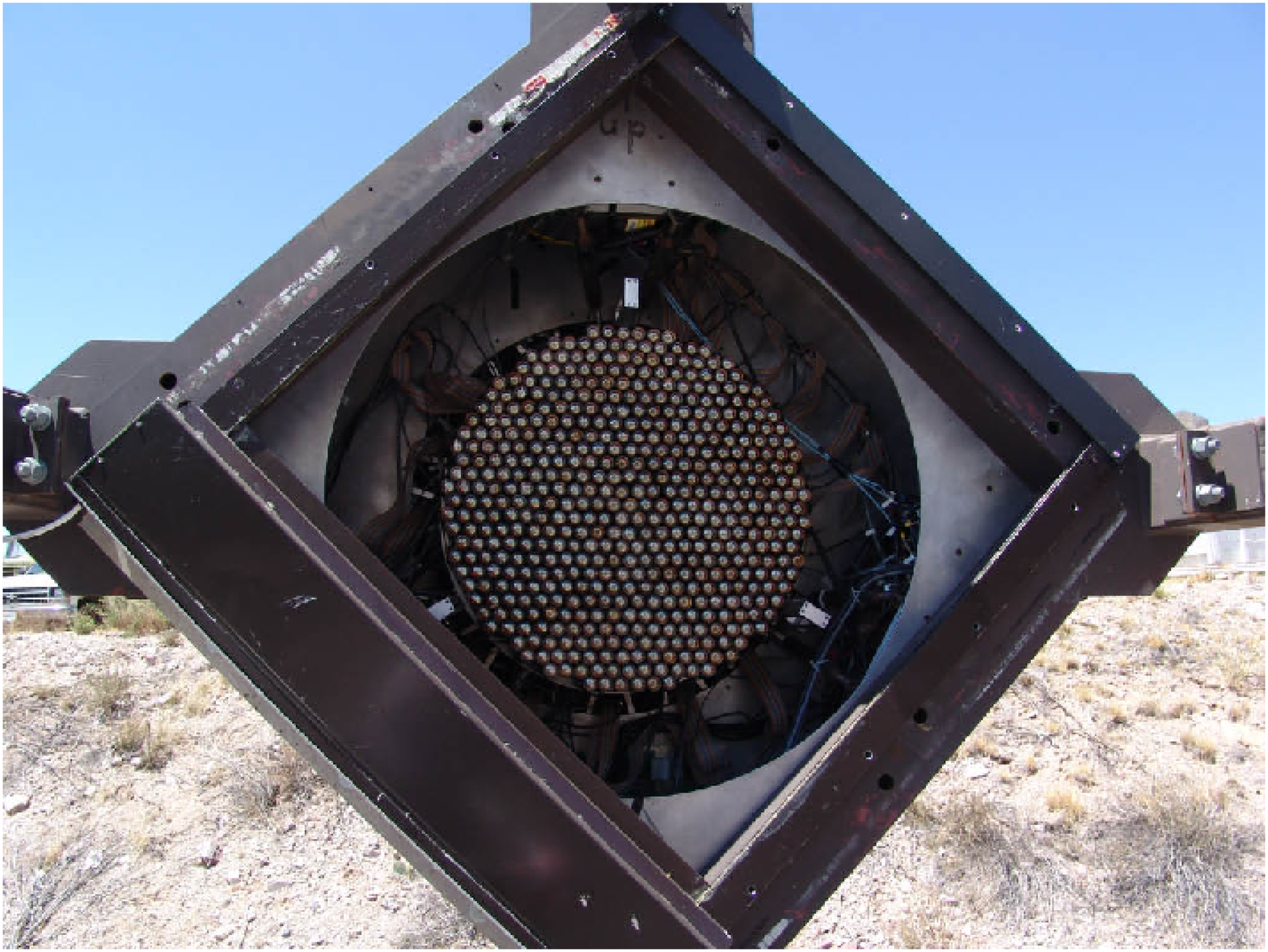,width=2.2in}\hspace{0.in}\psfig{file=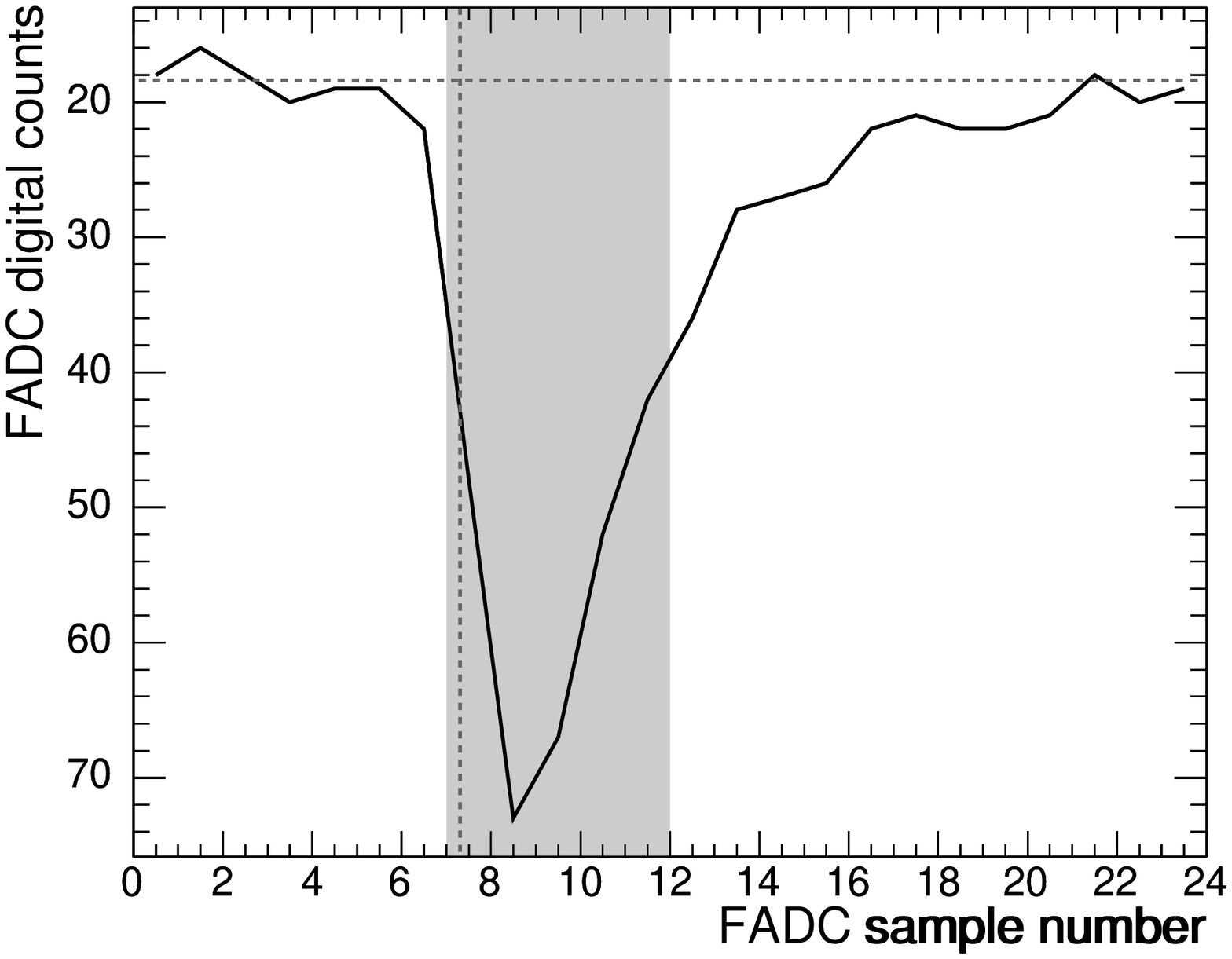,width=2.2in}
\end{center}
\caption{{\bf Left:} The 499 pixel imaging camera. The focus box is $1.8\U{m}$ square. Light cones have now been installed. {\bf Right:} A FADC trace resulting from Cherenkov light in a single PMT. The shaded area indicates a $10\U{ns}$ (5 FADC samples) integration window.}
\label{camera}
\end{figure}

The imaging camera, shown in Figure~\ref{camera}, consists of 499, $2.86\U{cm}$ diameter PMT pixels (Photonis XP2970/02). 
The pixel spacing corresponds to an angular separation of $0.15^{\circ}$, giving a total field-of-view of $\sim3.5^{\circ}$ diameter. 
Reflecting light cones have now been installed on the camera face and increase the overall photon collection efficiency by $\sim30\%$. 
The PMTs are operated at a gain of $\sim2\times10^{5}$, giving typical anode currents of $3\U{\mu A}$ (for dark fields) to $6\U{\mu A}$ (for bright fields). \
Signals from the PMTs are amplified by a high-bandwidth pre-amplifier integrated into the PMT base mounts and then sent via coaxial cable to the telescope trigger and data acquisition electronics housed in the control room.
The trigger electronics have two levels; each channel is equipped with a constant fraction discriminator (CFD) which produces an output logic pulse of programmable width (typically $10\U{ns}$) when the discriminator threshold is crossed\cite{Hall03}.
 These signals are then passed to a topological trigger system, similar to that used successfully on the Whipple 10m telescope \cite{Bradbury02}, which is used to detect pre-programmed patterns of triggered pixels in the camera (for example, any three adjacent pixels).
The topological trigger system greatly reduces the rate of triggers due to random fluctuations of the night sky background light and enabled us to operate the first telescope alone with a CFD threshold of $\sim7\U{photoelectrons}$ and a cosmic ray trigger rate of $\sim200\U{Hz}$ at high elevation.

In the event of a trigger, each PMT signal is digitized by a custom built, 8-bit Flash-ADC system, sampling at 500MHz with a memory depth of $32\U{\mu s}$ \cite{Buckley03}. 
The FADC readout window and size are programmable; Figure~\ref{camera} shows a typical FADC trace with a readout window of $48\U{ns}$ (24 FADC samples). The use of FADCs, as opposed to simple charge integrating ADCs, allows us to investigate various digital signal processing techniques to improve the signal/noise ratio in each channel, as well as providing information about the time structure of the Cherenkov image in the camera, which may be useful as an additional tool for shower parameter reconstruction and gamma-hadron discrimination \cite{Holder05}.

\vspace{-0.3in}
\subsection{Data Analysis and Results}
Prior to analysis, the signal channels must be calibrated. Relative (pixel-to-pixel) gain calibration is achieved by uniformly illuminating the camera face with a $\sim4\U{ns}$ pulse of laser light via an opal diffuser.
Electronic pedestal offsets are measured by force-triggering the data acquisition at a rate of $3\U{Hz}$ during observations.
Absolute calibration, which is necessary to understand the overall collection efficiency of the telescope, is achieved by measurements of the camera response to single photoelectron level signals \cite{Holder06} and to the Cherenkov light produced by local muons.

\begin{figure}
\begin{center}
\psfig{file=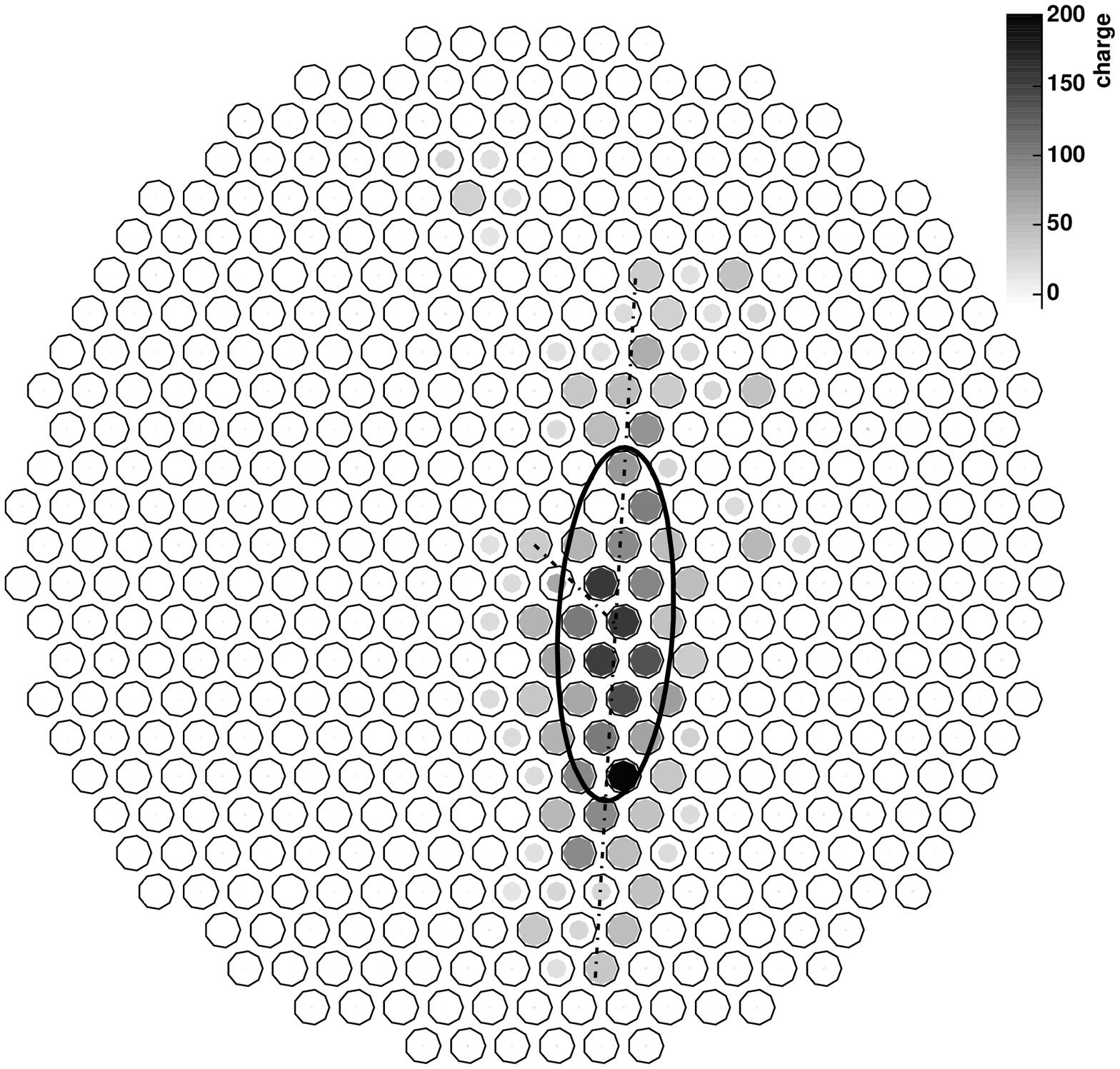,width=2.2in}\hspace{0.in}\psfig{file=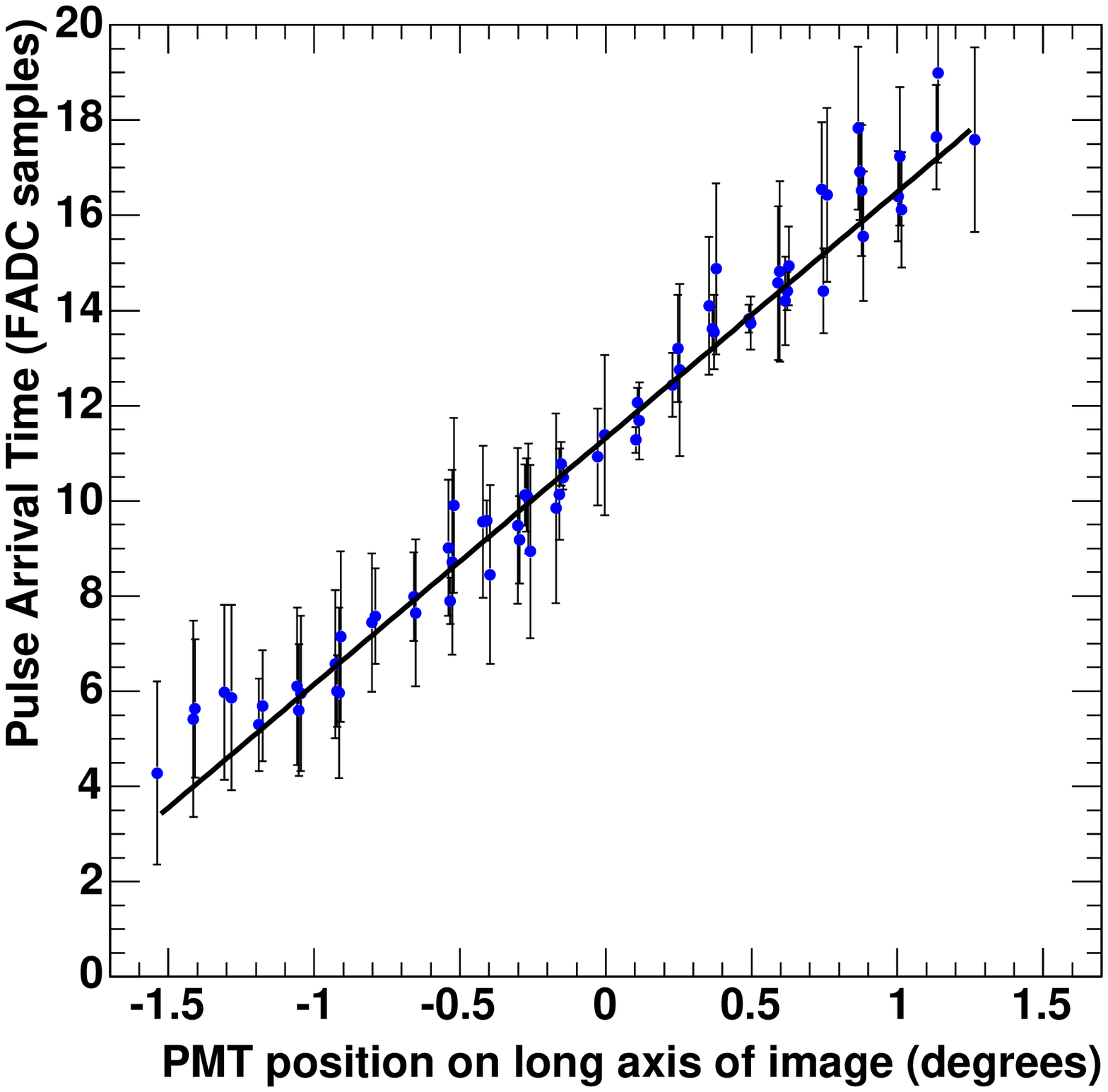,width=2.2in}
\end{center}
\caption{{\bf Left:} The charge distribution across the camera for a cosmic ray event. {\bf Right:} The pulse arrival time distribution along the long axis of the ellipse for the cosmic ray image on the left.}
\label{event}
\end{figure}

Following calibration, images are cleaned using a two-pass method.
In the first pass, a wide ($20\U{ns}$) integration window is applied to each FADC trace in order to calculate the integrated charge and the pulse arrival time.
PMTs with a clear signal are selected and the resulting image parameterised with a second moment analysis, the results of which can be described by an ellipse as shown in Figure~\ref{event}. 
Using the FADC information, the time gradient across the image can also be measured, and this information is used in a second pass over the image in order to position a shorter ($10\U{ns}$) charge integration window according to the expected position of the pulse.

\begin{figure}
\begin{center}
\psfig{file=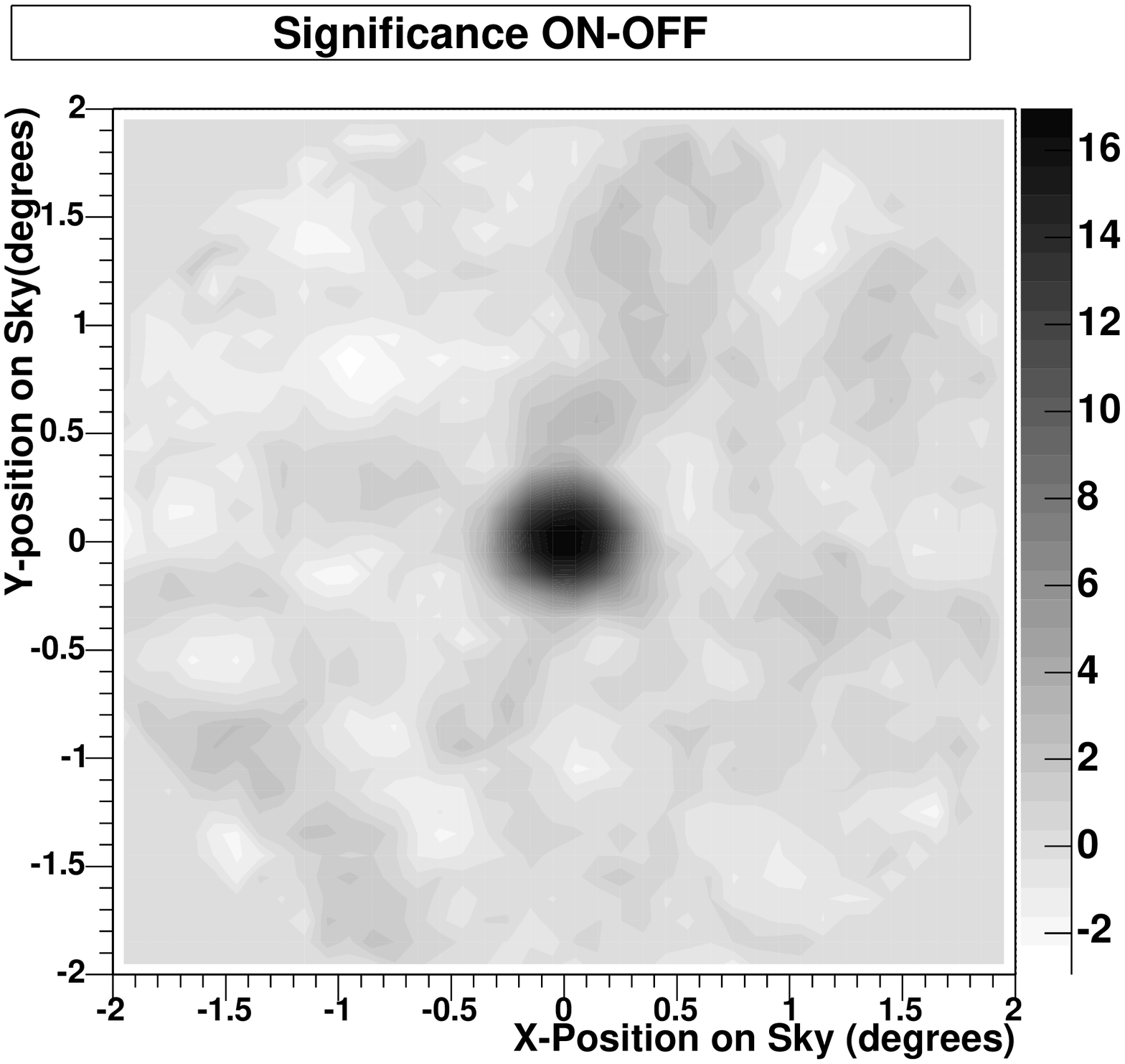,width=2.2in}\hspace{0.in}\psfig{file=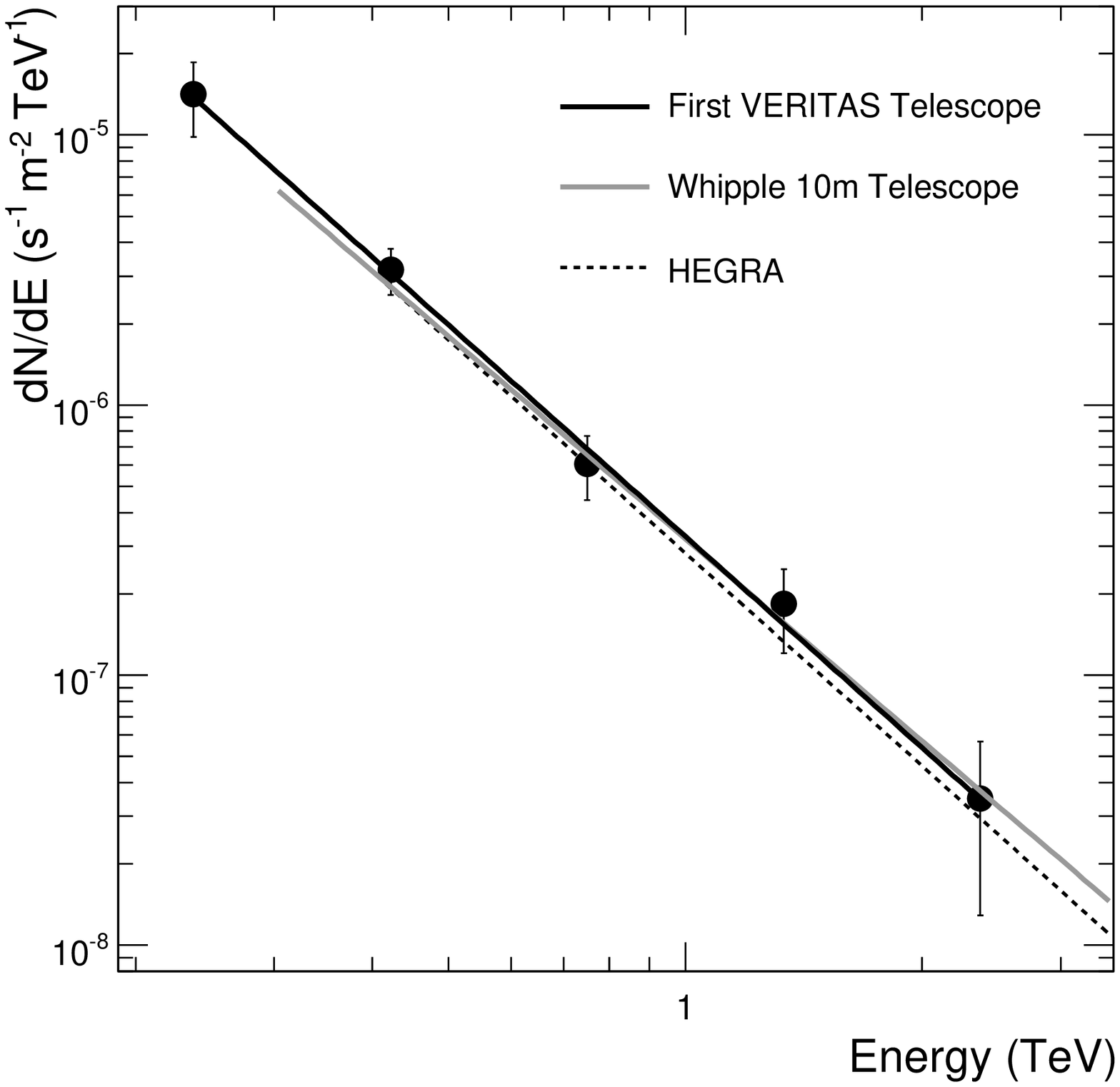,width=2.2in}
\end{center}
\caption{{\bf Left:} The map of reconstructed source position for observations of the Crab Nebula. The source was located at the centre of the camera. {\bf Right:} The energy spectrum of gamma-ray emission from the Crab Nebula compared with earlier results}
\label{crab}
\end{figure}

The standard candle of ground-based gamma-ray astronomy is the Crab Nebula, and observations of this object were made in order to test the performance of the first VERITAS telescope. Figure~\ref{crab} shows the map of the reconstructed source position for a 3.9 hour exposure to the Crab showing the source at the centre of the field-of-view. 
The single telescope sensitivity to the Crab was $\sim10\sigma/\sqrt{\mathrm{hour}}$. 
Also shown in Figure~\ref{crab} is the reconstructed energy spectrum of the gamma-ray flux from the Crab Nebula, as compared to earlier results.
A power-law fit to the data points gives a spectral index of $2.6\pm 0.3$ and a
differential flux at $1\U{TeV}$ of $(3.26\pm0.9)\cdot10^{-7}$
m$^{-2}$s$^{-1}$TeV$^{-1}$ (statistical errors only).
The energy reconstruction relies heavily on input from Monte Carlo simulations and so the Crab spectrum reconstruction (along with image parameter comparisons \cite{Maier05}) provides a confirmation that the telecope efficiency and operating parameters are well modelled.

\section{The Stereo System}

The second telescope was installed on the same site and began initial operations in January 2006. 
For the first few months the two telescopes operated independently and coincident events were identified offline using the GPS timestamps. 
In March 2006 a hardware array trigger system was installed which corrects the single telescope triggers for changing time delays due to the source movement across the sky.
The time-corrected signals are passed to a coincidence unit, and only events which trigger both telescopes within a coincidence time window of $100\U{ns}$ are read out. 

\begin{figure}
\begin{center}
\psfig{file=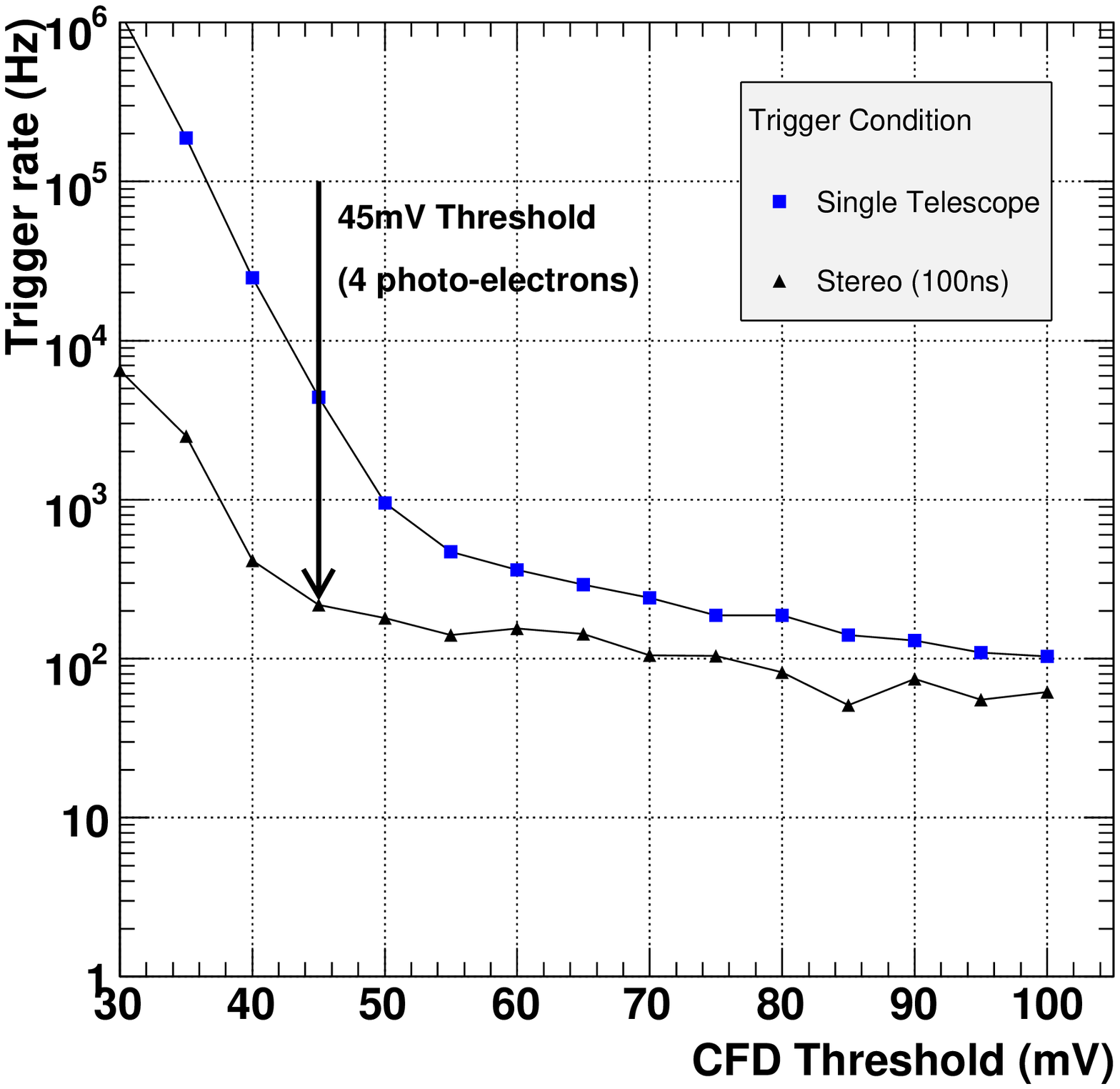,width=2.2in}\hspace{0.in}\psfig{file=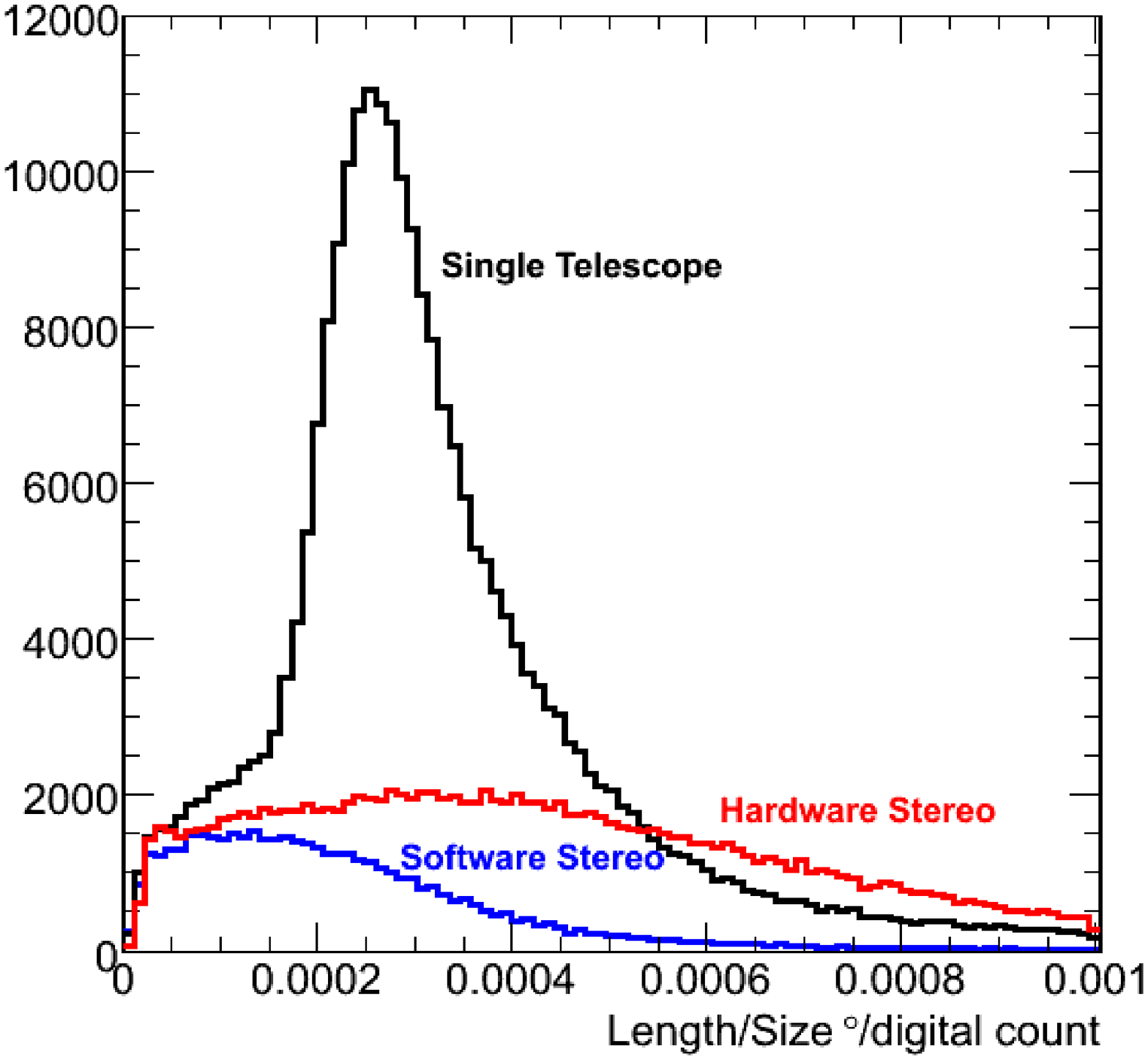,width=2.2in}
\end{center}
\caption{{\bf Left:} The trigger rate as a function of discriminator threshold for a single telescope and with the two telescope hardware trigger requirement. {\bf Right:} \textit {length/size} distributions. the single telescope histogram includes all events which trigger the telescope; the distinctive peak is due to local muons. The ``software stereo'' curve shows those events which remain after requiring that the events in both telescopes have a matching GPS timestamp within a $10\U{\mu s}$ window. The ``hardware stereo'' curve shows the same with the two-telescope hardware trigger operating.  }
\label{stereotrig}
\end{figure}

The stereo trigger serves two purposes: it allows us to run the telescopes at a lower discriminator threshold without being overwhelmed by night sky background triggers, and it removes events due to muons with impact parameters close to the individual telescopes, which provide an irreducible background for gamma-ray observations with a single telescope.
Figure~\ref{stereotrig} shows both of these effects; the left-hand plot illustrates that with the hardware array trigger requirement we are able to operate the system with single telescope discriminator settings equivalent to $\sim4\U{photoelectrons}$.
On the right we show the ratio of image \textit{length} to \textit{size} (integrated charge over the whole image).
Events due to local muons are distinguished by their relatively constant brightness per unit length, hence they display a distinctive peak in the single telescope histogram of $length/size$.
When the two-telescope requirement is applied, either in hardware or in software by matching event timestamps, this peak disappears, indicating that the muon events have been removed. 
Note also that the hardware requirement produces a distribution with many more small size events, as the triggering threshold is significantly lower in the hardware stereo as compared to the software stereo case.

The two telescope system was used to observe a selection of known gamma-ray sources from March to July 2006. The TeV blazar Markarian 421 was particularly active during this period and provided a high statistics sample of TeV gamma-rays with which to test the stereo system. Figure~\ref{stereo} shows the distribution of $\theta^2$ - the squared angular distance from the source location - for Mrk 421 observations. A strong source is evident.
 
\vspace{-0.0in}
\section{Conclusion}
\begin{figure}
\begin{center}
\psfig{file=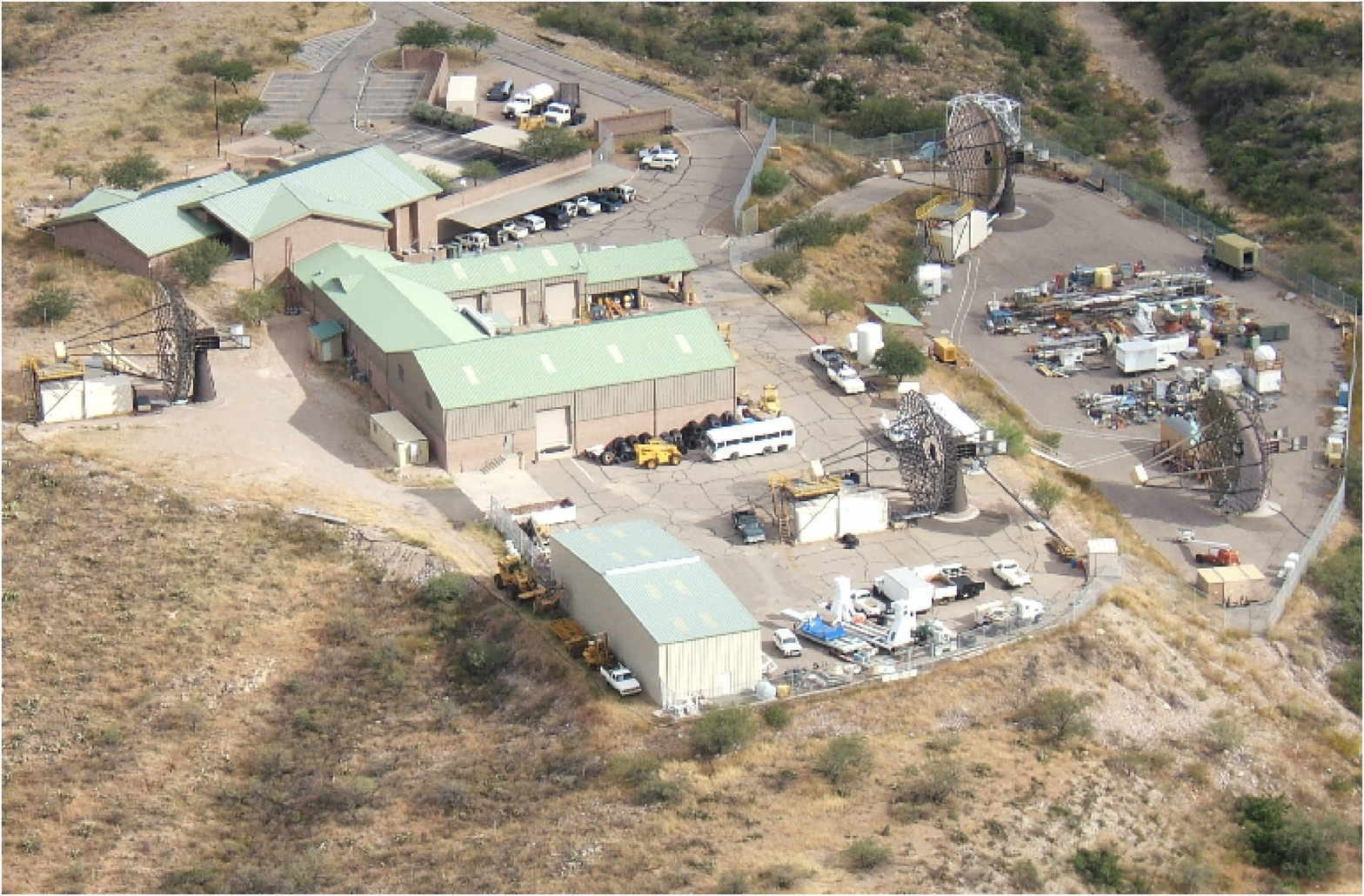,width=2.5in}\hspace{0.0in}\psfig{file=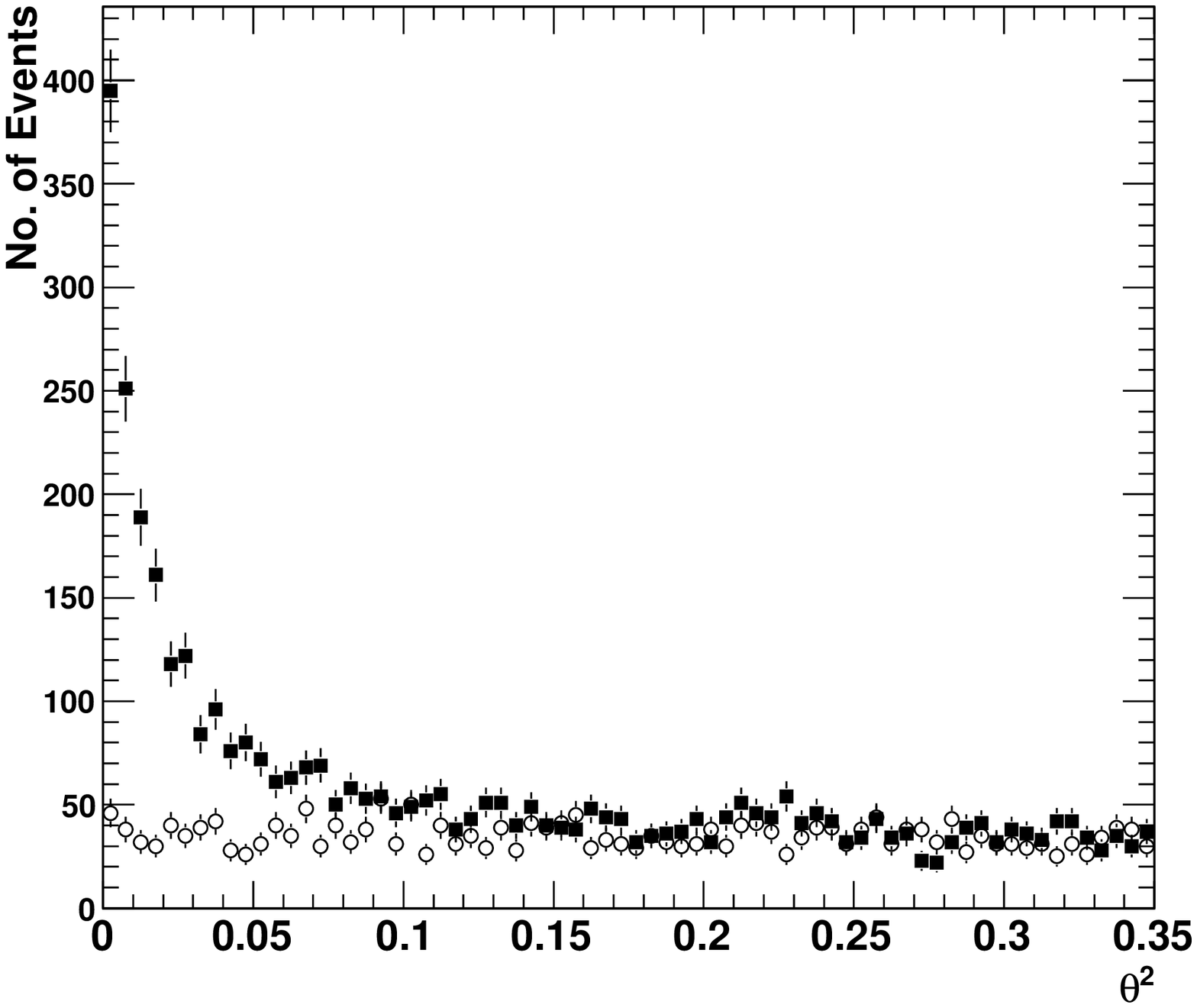,width=2.0in}
\end{center}
\caption{{\bf Left:} The four-telescope VERITAS array. {\bf Right:} The $\theta^2$ distribution for stereo observations of Markarian 421. Filled squares are observations on source, open circles are observations of an off-source control region.  }
\label{stereo}
\end{figure}

Observations made with the first VERITAS telescope, operating since early 2005, have enabled us to verify that all technical specifications have been. met. A number of previously known gamma-ray sources, including the Crab Nebula and Mrk~421 have been detected, and the single telescope simulations have been verified by comparison with real data. Initial stereo observations have shown that the hardware array trigger system is working as expected, and that the local muon background has been removed.
The construction of VERITAS is nearing completion. Figure~\ref{stereo} shows all four telescope structures assembled on site. Operations with the full array are scheduled to commence in January 2007, with a science program involving observations of supernova remnants, active galactic nuclei, and a survey of the galactic plane. 

\vspace{-0.3in}
\section{Acknowledgments}
Supported by grants from the U.S.~Department of Energy, the
U.S. National Science Foundation,
the Smithsonian Institution, by NSERC in Canada, by Science Foundation Ireland and by PPARC in the UK.

\end{document}